\documentclass[11pt,a4paper]{article}

\usepackage{amsmath,amssymb}
\usepackage{epsfig,graphicx}
\usepackage{subfigure}
\usepackage{graphicx}
\usepackage{rotating}
\usepackage{cancel}
\usepackage{bm}
\usepackage{color}
\usepackage{comment}
\usepackage{cite}
\usepackage{psfrag}

\newcommand{\wbase}{{\rm w}_{\rm base}}

\newcommand{\muu}{m}
\newcommand{\chie}{\chi_{\rm eff}}
\newcommand{\ve}[2]{\left(
\begin{array}{c}
 #1\\
#2
\end{array}
\right)
}

\renewcommand\({\left(}
\renewcommand\){\right)}
\renewcommand\[{\left[}

\newcommand{\J}[1]{\color{magenta} #1 \color{black}}

\newcommand{\exclude}[1]{}

\def\beq{\begin{equation}}
\def\eeq{\end{equation}}

\begin{document}
\numberwithin{equation}{section}
\title{{\normalsize  \mbox{}\hfill MPP-2013-223}\\
\vspace{2.5cm} 
\Large{\textbf{From Resonant to Broadband Searches\\ for WISPy Cold Dark Matter 
\vspace{0.5cm}}}}

\author{Joerg Jaeckel$^{1}$, and Javier Redondo$^{2,3}$\\[2ex]
\small{\em $^1$Institut f\"ur theoretische Physik, Universit\"at Heidelberg, Philosophenweg 16, 69120 Heidelberg, Germany}\\[0.5ex]
\small{\em $^2$Arnold Sommerfeld Center, Ludwig-Maximilians-Universit\"at, Munich, Germany}\\[0.5ex]
\small{\em $^3$Max-Planck-Institut f\"ur Physik, Munich, Germany}\\[0.5ex]
}

\maketitle

\begin{abstract}
\noindent
Dark matter may consist of light, very weakly interacting bosons, produced non-thermally in the early Universe.
Prominent examples of such very weakly interacting slim particles (WISPs) are axions and hidden photons.
Direct detection experiments for such particles are based on the conversion of these particles into photons. 
This can be done in resonant cavities, featuring a resonant enhancement, or by using suitably shaped reflecting surfaces that allow for broadband searches. In this note we want to elucidate the relation between the two setups and study the transition from resonant to broadband searches. This then allows to determine the sensitivity of off-resonance cavity searches for cavities much larger than the wavelength of the generated photons.\\

\end{abstract}

\newpage

\section{Introduction}
By now we have convincing observational evidence that roughly 27\% of the total energy/mass is in the form of 
dark matter (DM)~\cite{Ade:2013zuv}.
Nevertheless final confirmation of its existence and determination of its properties can only arise from direct detection.

At the moment our knowledge about the detailed particle physics properties of dark matter is very limited. Accordingly we still have viable candidates with very different properties. For example we have thermally produced weakly interacting massive particles (WIMPs) with large masses in the GeV to TeV range and interactions of similar strength to the weak interactions (see~\cite{Bertone:2004pz} for a review).
A typical realization being, for example, the (fermionic) neutralino in supersymmetry.
On the other hand we could also have very light and very weakly interacting non-thermally produced bosons. Prominent examples of such WISPy (for weakly interacting slim particle) dark matter~\cite{Preskill:1982cy,Abbott:1982af,Dine:1982ah,Sikivie:1982qv,Hiramatsu:2010yn,Piazza:2010ye,Nelson:2011sf,Arias:2012az} are axions and axion-like particles~\cite{Peccei:1977hh,Weinberg:1977ma,Wilczek:1977pj,Svrcek:2006yi
} as well as hidden photons~\cite{Dienes:1996zr} (see~\cite{Jaeckel:2010ni} for a review on general WISPs and~\cite{Jaeckel:2013ija} specifically for hidden-photons).

In this note we will focus on WISPs and in particular on the detection of axions, axion-like particles and hidden photons.
It is clear that such light particles require fundamentally different detection strategies than WIMPs.
For the direct detection of axions, axion-like particles and hidden photons two strategies have been suggested~\cite{Sikivie:1983ip,Horns:2012jf} (both may actually even allow for directional detection~\cite{Irastorza:2012jq,Jaeckel:2013sqa}). Both are based on the {\emph{conversion}} of DM WISPs into photons.
For axions and axion-like particles this conversion can occur via their two photon coupling in presence of a magnetic field
and for hidden photons it occurs via kinetic mixing without the need for a magnetic field.

For simplicity we will focus on hidden photons in the following. But, as discussed in~\cite{Horns:2012jf}, the case for axions is completely analogous.

The first strategy for converting DM WISPs into photons is a so-called cavity haloscope~\cite{Sikivie:1983ip}. Here the
conversion is enhanced by using a resonant cavity in which one resonant frequency is chosen to be equal to the mass of the DM particle.
This leads to a resonant amplification by the quality factor of the resonator. This approach is pursued in the search of axions by the ADMX collaboration~\cite{Asztalos:2003px} with additional experiments in planning. A clear advantage is the increased signal. However, this comes at the expense of having to scan through all the possible masses by tuning the cavity in small steps to achieve resonance for different masses (we do not know the  mass of DM particles!). 
The second approach, a dish antenna search, is broadband and uses the fact that reflecting surfaces convert DM WISPs into photons. If these surfaces are suitably shaped the signal can be concentrated in a small region where one can then place the detector~\cite{Horns:2012jf}. 

The picture usually employed to understand conversion in a cavity is that the WISP field provides a source for the photon field, which drives the cavity. This source is present everywhere in the cavity and accordingly conversion is a volume effect. 
Indeed, the usual formula gives the power output being proportional to the volume (times a geometric factor).
In contrast, the dish antenna approach is formulated in terms of a conversion on the surface. From this follows proportionality to the area. It is the main purpose of this note to connect these two pictures and demonstrate that they are just two different limits of the same physical effect.

This task might seem only of academic interest, but also serves a practical purpose. From the cavity viewpoint it naively seems advantageous to increase the volume. 
One could perhaps even give up resonant enhancement and do a broadband search with a very large cavity.
However, we point out that for frequencies large compared to the size of the cavity the emission is proportional to the area and not the volume, limiting the sensitivity for this approach.

This note is structured as follows. In the following Sect.~\ref{surface} we recall the essential features of surface emission. We use this opportunity to go beyond the ideal reflector and discuss the general situation of emission from
boundaries between media with different optical properties. We then demonstrate in Sect.~\ref{onedim} how resonant enhancement arises in a one dimensional example. 
In Sect.~\ref{threedim} we then consider a simple three dimensional cavity and discuss the transition between the resonant and non-resonant cases. Based on this example we discuss searches with very large cavities in Sect.~\ref{benefits}. We summarize and conclude in Sect.~\ref{conclusions}.

\section{Surface Emission}\label{surface}
Let us start by recalling the relevant equations for the case of hidden photon dark matter. 
In a suitable field basis (cf.~\cite{Horns:2012jf}) the Lagrangian describing hidden photons (HPs henceforth) is given by,
\begin{equation}
\label{HPlagrangian}
\mathcal L= -\frac{1}4F_{\mu\nu}F^{\mu\nu}-\frac{1}4 X_{\mu\nu} X^{\mu\nu}+\frac{m^2}2 (X_\mu X^\mu-2\chi A_{\mu}X^{\mu}+\chi^2 A_{\mu}A^{\mu})+ J^\mu A_\mu , 
\end{equation}
where $A_\mu,X_\mu$ are the photon and HP fields with field strengths $F_{\mu\nu},X_{\mu\nu} $, $m$ is the HP mass and $\chi$ is a tiny kinetic mixing parameter. 
In the low energy limit, HPs do not couple directly to any particle of the standard model, and their sole interaction is through the small mixing $\chi$ with photons.

Accordingly the equation of motion for plane waves with frequency $\omega$ and momentum $k$ is, 
\begin{equation}
\left[(\omega^2-k^2)\left(
\begin{array}{cc}
1  & 0     \\
0 &  1   \\  
\end{array}
\right)-
M^2
\left(
\begin{array}{cc}
1  & 0     \\
0 &  0   \\  
\end{array}
\right)
-
m^2\left(
\begin{array}{cc}
\chi^2 & -\chi     \\
 -\chi &  1    \\  
\end{array}
\right)
\right]
\left(
\begin{array}{c}
\mathbf{A}\\
\mathbf{X}
\end{array}
\right)=
\left(\begin{array}{c}
0\\
0
\end{array}
\right).
\end{equation}
Here we have used that  for small velocities $v=k/\omega \sim 10^{-3}$ of the dark matter particles we can choose a gauge such that $X^{0}\approx A^{0}\approx 0$. The corrections to this approximation are of order $v^{2}$~\cite{Jaeckel:2013sqa}.
Moreover, we have included refraction and absorption for the photons in the term 
\begin{equation}
M^2\equiv\omega^2(1-n^2) \equiv m_\gamma^2-i \omega \Gamma , 
\end{equation}
where $n$ is the complex index of refraction, $m_\gamma^2$ is the photon refractive mass and $\Gamma$ the absorption coefficient. (Throughout this paper we use natural units where $\hbar=c=1$.)

The two right moving solutions to this equation correspond to photon-like and HP-like 
\begin{eqnarray}
\left(
\begin{array}{c}
\mathbf{A}\\
\mathbf{X}
\end{array}
\right)
\bigg|_{\gamma-\rm like} &=&
\mathbf{A}_{0}\left(
\begin{array}{c}
1\\
\chie
\end{array}\right)\exp(-i(\omega t-k x)) , 
\\
\left(
\begin{array}{c}
\mathbf{A}\\
\mathbf{X}
\end{array}
\right)
\bigg|_{\gamma^\prime-\rm like}&=&
\mathbf{X}_{0}\left(
\begin{array}{c}
-\chie\\
1
\end{array}\right)\exp(-i(\omega t-p x)).
\end{eqnarray}
where the effective mixing is   
\begin{equation}
\label{eq:effmixing}
\chie=\chi\frac{\muu^2}{\muu^2-M^2}, 
\end{equation}
and the corresponding wavenumbers are given by\footnote{Our convention requires $k$'s and $p$'s with positive imaginary part.}\footnote{At order $\chi$ the masses are not changed by the mixing.} 
\begin{eqnarray}
k=\sqrt{\omega^2-m^2_k}\quad ; \quad m_k^2 &=& M^2-\frac{\chi(\chi_{\rm eff}-\chi)}{1+\chi^2}\muu^2 , 				\\
p=\sqrt{\omega^2-m^2_{p}}\quad ; \quad m_p^2 &=& \muu^2+ \frac{\chi(\chi_{\rm eff}-\chi)}{1+\chi^2}\muu^2.
\end{eqnarray}
\J{The polarization vector $\mathbf{A}_0$ is transverse to the propagation direction, i.e. $\mathbf{A}_0\cdot \mathbf{\hat z}=0$, but $\mathbf{X}$ can have any arbitrary direction. }
In the following we will consider a situation where the hidden photon impinges normal to the boundary surface. Moreover, we will only consider those vectorial components that are parallel to the plane. This said we will drop the vector indices for the moment and simply write $X$ for $\mathbf{X}_{\parallel}$ and analog, $A$ for $\mathbf{A}_{\parallel}$.

\begin{figure}[t]
   \centering
   \includegraphics[width=8cm]{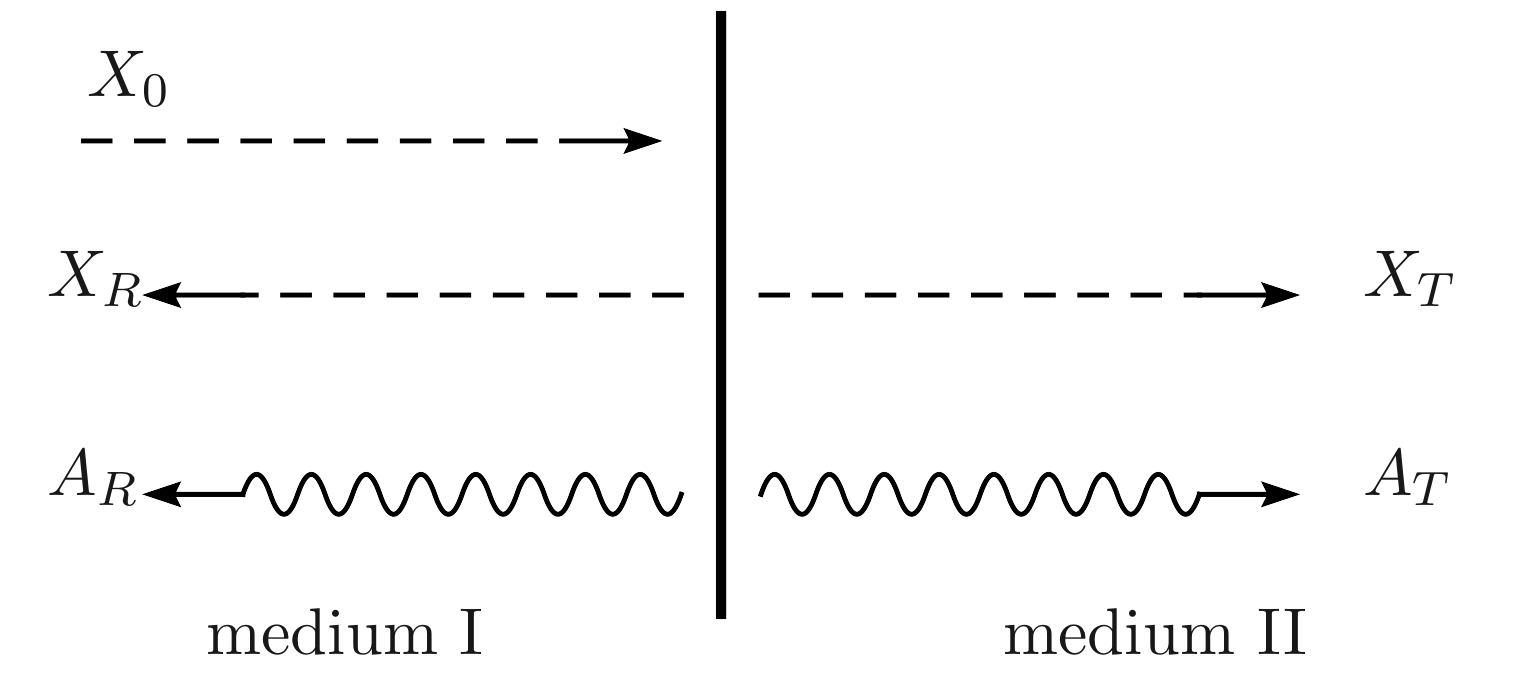} 
   \caption{Sketch of the various ingoing, reflected and transmitted hidden photon waves (dashed) and the transmitted and reflected photon waves (wavy) at the transition between two media.}
   \label{fig:sketch}
\end{figure}

Let us consider the passage of a HP-like state from a medium-I to a medium-II. (far away boundaries).
The situation is sketched in Fig.~\ref{fig:sketch}.  
The solution to the equations of motion in each of the media is of course a superposition of plane-waves 
\begin{eqnarray}
{\rm 1}&=&
X_0 \ve{-\chi_1}{1} 	\exp(i p_1 x)
+X_R \ve{-\chi_1}{1} \exp(-i p_1 x)
+A_R\ve{1}{\chi_1} 	\exp(-i k_1 x)			\\
{\rm 2}&=&
  X_T \ve{-\chi_2}{1} \exp(i p_2 x)
+A_T\ve{1}{\chi_2} 	\exp(i k_2 x)			\\
\end{eqnarray}
and we can obtain the amplitudes of the different photon-like or HP-like waves by imposing the continuity of the solution and its first derivative at the boundary $x=0$, 
\begin{eqnarray}
\frac{X_T}{X_0}&=& \frac{2 p_1}{p_2+p_1}, 					\\
\frac{X_R}{X_0}&=& \frac{p_1-p_2}{p_2+p_1}	, 				\\
\frac{A_T}{X_0}&=& -(\chi_1-\chi_2)\frac{2 p_1}{p_2+p_1}\frac{k_1+p_2}{k_1+k_2}, 	\\
\frac{A_R}{X_0}&=& (\chi_1-\chi_2)\frac{2 p_1}{p_2+p_1}\frac{k_2-p_2}{k_1+k_2}, 		
\end{eqnarray}
at first order in $\chi_{1}, \chi_2$. 
These are the generalization of the Fresnel equations for the photon-HP system at normal incidence. 

We can simplify these expressions in the physical case at work by considering that HPs conforming the local dark matter should have velocities of order $v\sim 10^{-3}=300\,{\rm km/s}$. 
This corresponds to $p_1\sim p_2\sim v \muu \ll \muu$ with tiny corrections of order $\chi^2$ that we will neglect. 
Also, typically we can neglect $p_2$ compared to $k$'s because the latter will be of order $\omega$. We then find
\begin{eqnarray}
\frac{X_T}{X_0}&\simeq&   1					\\
\frac{X_R}{X_0}&\simeq&   0					\\
\frac{A_T}{X_0}&\simeq& -(\chi_1-\chi_2)\frac{k_1}{k_1+k_2} \\
\frac{A_R}{X_0}&\simeq& (\chi_1-\chi_2)\frac{k_2}{k_1+k_2}		
\end{eqnarray}
which shows that photon-like waves are generated to compensate the change of the photon-component of the DM eigenstate across the boundary. 
Importantly, these waves are proportional to the change in the effective mixing from medium 1 to 2, $\chi_1-\chi_2$. 
In order to have a sizable photon emitted from the boundary, the effective mixing has to change by an ${\cal O}(1)$ amount. Using Eq.~\eqref{eq:effmixing} and taking into account that $\omega\approx m$ we find that this only happens if the complex index of refraction changes by an ${\cal O}(1)$ amount, as
\begin{eqnarray}
\chi_1-\chi_2 = \chi\(\frac{1}{n_1^2}-\frac{1}{n_2^2}\)	\ . 
\end{eqnarray}

\subsection{Examples}
\subsubsection*{Exiting a mirror}
We consider the situation depicted in Fig.~\ref{fig:sketch} when the LHS medium is a reflector and the RHS medium is vacuum ($ k_2 = \omega,\chi_2=\chi$).  
A good reflector is defined such that its reflection coefficient $|(\omega-k_1)/(\omega+k_1)|^2$ is close to one. This requires $|k_1|\gg \omega$, i.e. $|n_1|\gg 1$.  In such a case, the effective mixing are $\chi_1\ll \chi$ and the photonic wave generated after the mirror is
\begin{equation}
\frac{A_T}{X_0} \simeq -(\chi_1-\chi_2)\frac{k_1}{k_1+k_2}\to \chi . 
\end{equation}

Such good reflectors can be realized in a number of ways. 
If we think of a metallic surface, at sufficiently low frequencies the complex index of refraction is 
$n^2=1- i \sigma_{\rm DC}/\omega$ and dominated by the DC conductivity ($\sigma_{\rm DC}$).  
At sufficiently low frequencies $n$ becomes then much larger than one and imaginary, i.e. 
we have $|k_1|\gg \omega$. At higher frequencies, in the diffusive (Hagen-Rubens) regime, 
the index of refraction is given by the plasma frequency $n^2= 1-(\omega_P/\omega)^2$ and again $|n|$ can be 
larger than one for frequencies smaller than $\omega_P$. This last consideration also applies for plasmas, although they promise few practical applications\footnote{For instance, it is well known that the ionosphere reflects electromagnetic waves in the radio regime. Actually, it is amusing to consider very low mass HPs causing the emission of radiation out of Earth's ionosphere. }. 

\subsubsection*{Entering a mirror}
In this case, the RHS medium is the reflector and the LHS is vacuum. The roles of $1$ and $2$ are exchanged with respect to the previous case. The reflected wave is 
\begin{equation}
\frac{A_R}{X_0}\simeq (\chi_1-\chi_2)\frac{k_2}{k_1^*+k_2}\to \chi 
\end{equation}

\subsection*{Radiation from dielectrics}

It is worth noting that the radiation of photons does not only occur from perfectly reflecting surfaces. 
We expect radiation from any interface where the index of refraction changes by ${\cal O}(1)$. 
Let consider the interface between air and a flint glass and $\muu$ in the near infrared ($n_{\rm flint}\sim 1.6$).  
The radiated waves from a HP entering into the glass will be 
\begin{equation}
\frac{A_T}{X_0} \simeq -0.23 \chi \quad ; \quad 
\frac{A_R}{X_0} \simeq 0.38 \chi .
\end{equation}
which can be still sizable in absolute value (depending on $\chi$) despite the small reflection coefficient $\sim 0.05$. 

\subsection*{Dielectric mirrors} 

A dielectric mirror is made of a sequence of layers of dielectric alternating high and low index of refraction. 
Unfortunately, the simple formulas above are not very useful to evaluate what is the response of such a complex system to the passage of a HP. One can find the reflected and transmitted photonic waves after generalizing the transfer-matrix formalism from two (right and left photonic waves) to four components (right and left HP-like and photon-like waves). We have done so but find it very cumbersome and a bit sterile when it comes to results. 
Instead of going into details let us comment on one important example: a $\lambda/4$ dielectric mirror designed for normal incidence. 
This type of mirrors consists of a sequence of $2N$ alternated layers of high and low index of refraction ($n_h,n_l$). 
The thickness of the high and low index of refraction layers ($d_h,d_l$) are such than the waves undergo one quarter of an oscillation, i.e. $n_h d_h=\lambda/4$ where $\lambda$ is the vacuum wavelength  (the expression for $d_l$ is analogous). 
Such mirrors are specifically designed for a certain wavelength $\lambda=2\pi/\omega$.
With this construction, the reflected waves from surfaces between high and low and those between low and high, which have a phase difference of -1, interfere constructively. 
One can have a reflectivity very close to 1 by stacking enough repetitions of high-and-low layers even if the reflectivity of each of the contact surfaces is small.  

\begin{figure}[t]
\centering
   \includegraphics[width=12cm]{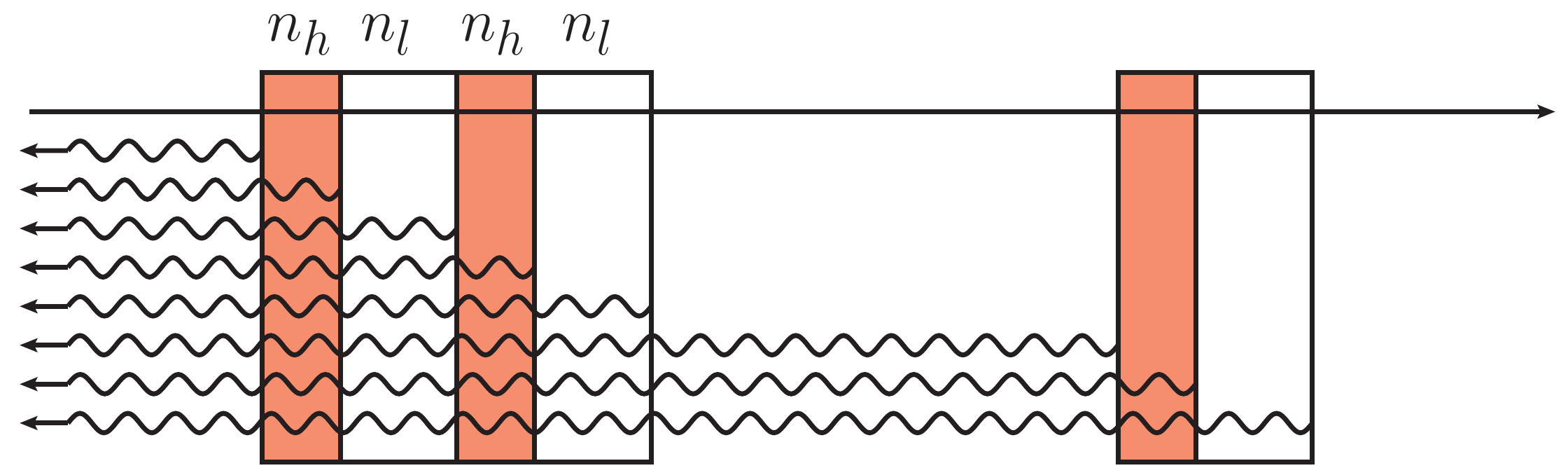} 
   \caption{Simplified scheme for the photon wave created by the passing of a HP through a dielectric mirror of low reflectivity surfaces.}
   \label{fig:dielectricmirror}
\end{figure}

When a HP goes through one of these mirrors all the $2N+1$ surfaces radiate coherently in both directions. 
The radiated waves are transmitted and reflected in all the surfaces leading to a very complex system in general. 
For the sake of illustration let us assume that the reflectivity of each of the contact surfaces is very small and focus on the reflected wave when an HP enters from vacuum into the $\lambda/4$ mirror. The situation is depicted in Fig.~\ref{fig:dielectricmirror}.  
The dominant rays will come from each of the contact-surfaces. There are four types of surfaces: vacuum-H, H-L, L-H and L-vacuum. Each of these waves is retarded by $\lambda/4$ with respect to the previous one so we have
\begin{eqnarray}
\label{eq:dielectricexercise}
A^{\lambda/4}_{R,\rm full}&\approx& \nonumber
A_{R,\rm vacuum-H}+\(i A_{R,\rm H-L}+ i^2 A_{R,\rm L-H}\)+
\(i^3 A_{R,\rm H-L}+ i^4 A_{R,\rm L-H}\) \\ 
&& + ... +i^{2N+1} A_{R,\rm L-vacuum} \\
&=& 
X_0\left\{
\begin{array}{l l}
\frac{(\chi-\chi_h)n_h}{1+n_h}-i \frac{\chi_l-\chi}{1+n_l}+i (\chi_h-\chi_l)\frac{n_l-in_h}{n_h+n_l} & (N\  \rm odd) \\
\frac{(\chi-\chi_h)n_h}{1+n_h}+i \frac{\chi_l-\chi}{1+n_l} & (N\  \rm even) \\
\end{array} 
\right.
\end{eqnarray}
We see that the emission from the first group of H-L layers cancels the emission from the second group, and so forth. If $N$ is even all the emission from the H-L and L-H surfaces cancels out and we are left with the emission from the first and last surface with a phase difference of $\pi/2$. If $N$ is even the emission from one group of H-L layers also survives. At the end, the reflected wave is of order $\chi$ 
and has a large phase. A similar situation happens for the transmitted wave. 
Thus one can use dielectric mirrors for DM detection experiments, but the analysis of the signal requires a dedicated an non-trivial calculation. 

The above exercise invites to find dielectric mirrors for which all the emitted waves are in phase, and the resulting reflected wave is amplified by the number of layers. 
At the view of \eqref{eq:dielectricexercise}, and noting that $A_{R,\rm H-L}$ and $A_{R,\rm L-H}$ have a minus sign difference, we realize that this can be achieved by building a set of dielectric layers just like the one discussed but with $\lambda/2$ thickness instead of $\lambda/4$. 
In such a case we have  
\begin{eqnarray}
\label{eq:dielectricexercise2}
A^{\lambda/2}_{R,\rm full}&\approx& \nonumber
A_{R,\rm vacuum-H}-A_{R,\rm H-L}+ A_{R,\rm L-H}-
A_{R,\rm H-L}+ A_{R,\rm L-H} + ... +(-1)^{2N+1} A_{R,\rm L-vacuum} \\
&\approx& 
N X_0\( A_{R,\rm L-H}- A_{R,\rm H-L}\)=N X_0\(\chi_l-\chi_h\)\frac{n_h-n_l}{n_h+n_l} \ , 
\end{eqnarray}
and the wave is enhanced by a factor ${\cal O}(N)$ (we have neglected the waves from the first and last surfaces). 
Doing the full calculation including multiple reflections with the transfer-matrix formalism we have 
obtained 
\begin{eqnarray}
\label{eq:dielectricexercise2full}
A^{\lambda/2}_{R,\rm full}&\approx& N X_0\(\chi_l-\chi_h\) . 
\end{eqnarray}

Such a sequence of $\lambda/2$ layers of alternating high and low index of refraction is of course not a mirror.  With such a phase difference the reflections in H-L and L-H surfaces cancel each other and one obtains a transmission filter. Using $\lambda/2$ dielectric filters we expect DM radiated waves with a power amplification\footnote{Coherence again restricts the maximal useable $N$ to be $n\sim 1/v\sim 10^3$ and the resonant power enhancement is $\sim N^2\sim 10^{6}$ as in the case of resonant cavities. $\sim N^2$.}
Since we don't know the mass of the DM HP, this discussion might sound a bit ludicrous. 
Barring dispersion, a $\lambda/4$ mirror behaves as a $\lambda/2$ filter for a wavelength twice as short as the design wavelength. Thus we should state that any sequence of alternating H-L dielectrics with equal optical path, i.e. $n_h d_h = n_ld_l$, emits photonic waves of amplitude $N X_0 (\chi_l-\chi_h)$ and (following arguments of symmetry) $N X_0 (\chi_h-\chi_l)$ out of its facets, \emph{ if the HP mass is $m=\pi /n_h d_h$}. 
The arguments given above also hold for $3\lambda/2$, etc.. so this resonant enhancement of the signal applies for masses in small intervals around 
\begin{equation}
m=\frac{\pi + 2\pi s}{n_l d_l} \quad ; \quad s\in \mathbb{Z} \ . 
\end{equation}

\section{Two facing mirrors: a one dimensional resonator}\label{onedim}

Let us consider a situation where a dark matter HP crosses a resonant cavity composed of two facing mirrors, whose reflecting surfaces are separated by a distance $L$. In this case, both mirrors emit photonic waves excited by the passing HP. 
These waves bounce back and forth between both mirrors and produce an interference pattern. 
The wave emitted by the LHS and RHS mirrors are originally 
\begin{equation}
A_T=\chi X_0 \exp\(i\omega x\)\quad ; \quad 	A_R=\chi X_0 \exp\(-i\omega x -i \delta_p\)
\end{equation}
where $\delta_p=p L=\sqrt{\omega^2-m_{\gamma'}^2}$ corresponds to the phase difference of the dark matter HP wave between the two mirrors. 
Every time a photon reaches a mirror, it produces a reflected wave of identical frequency, opposite phase and a slightly decreased amplitude due to finite transmission or absorption. 
After a round-trip the twice reflected wave is proportional to the original wave multiplied by a factor $r^2\exp(i 2 \omega L) $ where $r$ is the reflection coefficient and $2\omega L\equiv 2\delta$ is the round-trip phase. 
Summing over all bounces, the right-moving wave inside the cavity is 
\begin{eqnarray}
A(x) &=& \chi X_0 \exp(i\omega x) \(
 \sum_{b=0}^{b=\infty}\(r^2 \exp (2 i\delta)\)^b-r \exp(i (\delta-\delta_p))\sum_{b=0}^{b=\infty}\(r^2 \exp (2 i \delta)\)^b\) \\
&=& \chi X_0 \exp(i\omega x)  \frac{1-r \exp(i (\delta-\delta_p)}{1-r^2\exp(2i\delta)} .   
\end{eqnarray}
This is the type of solution we would obtain for an optical resonator \emph{that is fed from both sides coherently} (with a small phase difference given by $\delta_p$). Since this is not a very well known situation a closer look is in order.  

First we use $r=\sqrt{1-t^2}$, where $t\ll 1$ is the transmission coefficient 
\begin{equation}
A(x) \xrightarrow{t\ll1} \chi X_0 \exp(i\omega x)  \frac{1-\exp(i (\delta-\delta_p)+t^2/2}{1-\exp(2i\delta)+ t^2} .   
\end{equation}
where we have taken into account that factors of $t^2$ only matter when the $1-\exp()$ factors are closely tuned to zero, 
which correspond to $\exp()\sim 1$. 
The denominator is minimal when $2\delta = 2\pi n $ ($n\in \mathbb{Z}$), i.e. when the round-trip phase is an integer number of complete oscillations. This is completely analogous to a one-sided-fed resonator. There we have two types of modes according to the phase acquired by the wave after the cavity's length, $\delta$: even modes where $\exp ( i \delta)=1$ and odd modes where $\exp ( i\delta)=-1$. 

However, in our case, the second mirror is also emitting and this wave can interfere destructively or constructively. 
Taking $\delta_p=0$ for the moment, we see that the numerator cancels up to a factor $t$ for even modes, and it is 
$2$ for odd modes. Therefore:
\vspace{0.7cm}

\emph{Among all the resonances we have in a one-sided-fed resonator, only those where the phase is opposite in both mirrors, i.e. $\delta = \pi + 2\pi n$ can resonate when excited by WISPy dark matter. }
\vspace{0.1cm}

This conclusion holds to a good approximation when we include the effects of $\delta_p$. 
The reason is that dark matter waves have $p\ll \omega$ so that typically $\delta \gg \delta_p$. 
In the galaxy the typical dark matter velocity is $v= p/\omega\sim 10^{-3}$ so that only when we consider very high modes $n\sim {\cal O}(10^3)$ the combination $\exp(i(\delta-\delta_p))$ can differ significantly from $\exp(i\delta)$. As $\delta_p$ grows even modes get very slowly less suppressed and odd modes become more suprpessed. 
Eventually when $\delta_p=\pi$ even and odd-modes exchange their role and as $\delta_p$ grows the role of the resonant modes is alternatively taken by even or odd.  

In a one dimensional situation the square of amplitude of the outgoing photon wave is the emitted power per area. 
Neglecting $\delta_p$, the output power per area after the RHS mirror is thus approximately,  
\begin{equation}
\label{outputpower}
\frac{P_{\rm out,one\, side}}{{\rm Area}} \simeq  \chi^2\omega^2 X_0^2 \frac{1-\cos(\omega L)}{1-\cos\(2\omega L\)+t^4/2}t^2
\end{equation}
where the last $t^2$ is the power transmissivity of the mirror. 
The power exiting from the LHS will obviously be the same but the wave will have opposite phase. 

We can relate $|X_0|^2$ to the density of DM. Recalling that we used $\mathbf{X}=(\mathbf{X}_{||}+\mathbf{X}_\perp )e^{-i(\omega t + p x)}$ with $|\mathbf{X}_{||}|=X_0$ and neglecting small effects like kinetic energy and the small photon component, the energy density of dark matter is 
\begin{equation}
\rho_{\rm CDM} \approx \frac{1}{2} |\partial_t X|^2+\frac{1}{2} m^2 |X|^2 \approx m^2 |X|^2 = m^2 \frac{|X_0|^2}{\cos^2\theta} ,  
\end{equation}
where we have introduced the angle $\theta$ between a general $\mathbf{X}$ and the surface. The perpendicular component of an arbitrarily oriented $\mathbf{X}$ does not contribute to the radiation and so we have ignored it above. However, it contributes to the energy density and we should include it in this formula. 

The power emitted by both sides of the cavity is then 
\begin{equation}
\frac{P_{\rm out}}{{\rm Area}} \simeq  2 \chi^2 \rho_{\rm CDM}\cos^2(\theta) \frac{1-\cos(\omega L)}{1-\cos\(2\omega L\)+t^4/2}t^2 . 
\end{equation}

\subsection{Comparison with a resonator calculation}
Let us now compare this result with that of a standard cavity calculation~\cite{Sikivie:1983ip,Arias:2012az}.
On resonance (we will discuss the off-resonant case below) the cavity result gives,
\begin{equation}
\label{cavityout}
P_{out}=\kappa\chi^2 \omega \rho_{\rm CDM} QV{\mathcal{G}},
\end{equation}
where $\kappa$ is the coupling of the cavity. In our simple case, where all ``losses'' arise from photons leaving the cavity through the mirror and leaving to the detector $\kappa=1$.
The geometry factor is,
\begin{equation}
\label{geometry}
\mathcal G
= \frac{\left|\int d^{3}\mathbf{x}\,{\mathbf{A}}^{*\rm cav}(\mathbf{x})\cdot \hat{\mathbf{n}}\right|^2}{V\, \int d^3\, \mathbf{x} |\mathbf{A}^{\rm cav}(\mathbf{x})|^2},
\end{equation}
with,
\begin{equation}
\hat{\mathbf{n}}=\frac{\mathbf{X}}{|\mathbf{X}|}.
\end{equation}

Now the specific result depends on the direction of the hidden photon vector. If all hidden photons in the dark matter point in the same direction,
\begin{equation}
\mathbf{X}=\hat{\mathbf{n}}\frac{\sqrt{\rho_{\rm CDM}}}{m_{\gamma^{\prime}}}.
\end{equation}
For simplicity this is what we will assume in the following. The more general case can be recovered, by a suitable average 
over the distribution of the hidden photon directions.

In our one dimensional cavity the photon field of the cavity modes is parallel to the reflector plane. So the the scalar product in the geometry factor Eq.~\eqref{geometry} ensures that only the components parallel to the plane contribute.
 
We can now easily evaluate the geometry factor for our one dimensional cavity,
\begin{equation}
{\mathcal{G}}=\bigg\{\begin{array}{lcl}
\frac{8}{\pi^2n^2}\cos^{2}(\theta) & {\rm for} & n=odd\\
0 & {\rm for} & n=even,
\end{array}
\end{equation}
where $\theta$ is the angle between the hidden photon field and the reflector plane.

As in our calculation via surface emission we see that only the modes with an odd number of half-wavelength resonate.

Let us now compare the total output power on resonance.
For our Fabry-Perot type setup the quality factor is,
\begin{equation}
\label{fabryq}
Q=\frac{n\pi}{t^2}.
\end{equation}
The magnitude of the field in the direction parallel to the reflector is,
\begin{equation}
X_{0}=|\mathbf{X}_{\parallel}|=\cos(\theta)|\mathbf{X}|=\cos(\theta)\frac{\sqrt{\rho_{\rm CDM}}}{\omega},
\end{equation}
where we have used $m_{\gamma^{\prime}}\approx\omega$.

Putting everything together for our cavity,
\begin{equation}
\frac{P_{\rm out}}{\rm Area}=8\chi^2\rho_{\rm CDM}\cos^2(\theta)\frac{1}{t^2}, 
\end{equation}
which agrees with Eq.~\eqref{outputpower} when we are on resonance.

Importantly we note that while Eq.~\eqref{cavityout} naively suggests an increase of the output power with the volume and therefore with the length of the resonator this is actually not the case. The increase in volume is more than eaten up by a decrease of the geometry factor and it is only in combination with an increasing quality factor that we get a constant output power. This is then exactly in line with the expectation from surface emission.

\section{Power production and power density inside a large cavity}\label{threedim}
Let us continue with our discussion in \J{the} three dimensional case.
A technically important situation arises when we consider a vary large cavity. I.e. we consider a situation where the oscillation frequency of the hidden photons,
\begin{equation}
\omega=\sqrt{m^2+p^{2}}\approx m_{\gamma^{\prime}}\gg \frac{2\pi}{L_{\rm cavity}},
\end{equation}
where $L_{\rm cavity}$ is the typical spatial extent of the cavity.
In other words we consider a situation where the hidden photon dark matter will excite very high spatial modes inside 
the cavity.

\subsection{Surface emission}

A simple way to calculate the produced power is to simply consider (incoherent) emission from the surface as it was done in the calculation for dish antenna searches~\cite{Horns:2012jf}. We can then simply integrate the power emitted per area over the surface area.
Doing this we find that the emitted power to the inside of the cavity is given by,
\begin{equation}
\label{surfaceemission}
P_{\rm surface}=\langle \cos^{2}(\theta) \rangle_{\rm surface}\chi^2\rho_{\rm CDM} A_{\rm cavity}.
\end{equation}
The average in Eq.~\eqref{surfaceemission} is taken over the whole surface of the cavity.

Before we take a look from the viewpoint of a proper oscillator let us evaluate Eq.~\eqref{surfaceemission} for a simple cubic box cavity with side length $L$. For this simple configuration $\langle \cos^{2}(\theta)\rangle_{\rm surface}=2/3$ and we have
\begin{equation}
\label{cubic1}
P^{\rm cubic}_{\rm surface}=4 L^2 \chi^2 \rho_{\rm CDM}.
\end{equation}

\subsection{Cavity calculation}

Let us now take an approach from the cavity perspective.
A crucial feature is that we have a large number of narrowly spaced modes and we have to sum over all the relevant ones. 

\subsubsection{Power emitted from one mode}
Let us start by recalling the usual computation~\cite{Arias:2012az}, allowing it to be off-resonance.
The photon field ($\mathbf{A}$) inside the cavity can be expanded in terms of the cavity modes, 
\begin{equation} 
\mathbf{A}(\mathbf{x})
=\sum_i \alpha_i \mathbf{A}^{\rm cav}_i(\mathbf{x}), \ \ \ \ \  \int d^3\mathbf{x} |\mathbf{A}^{\rm cav}_i(\mathbf{x})|^2=C_i,
\end{equation}
with $C_i$ the normalisation coefficients.
Using this expansion and including losses in the cavity we obtain for the expansion coefficients,
\begin{equation}
\left(\frac{d^{2}}{dt}+\frac{\omega_{0}}{Q}\frac{d}{dt}+\omega^{2}_{0}\right) \alpha_{i}(t)=b_{i}\exp(-i\omega t),
\end{equation}
with $\omega_0$ the frequency of the cavity and $Q$ its quality factor.
The driving force $b_{i}$ can be written as
\begin{equation}
b_{i}=\frac{\chi m^2}{C_{i}}  
\int dV {\mathbf{A}}^{\star}_{i}(\mathbf{x})\cdot\mathbf{X}(\mathbf{x}).
\end{equation}

The asymptotic solution for the cavity coefficients is then,
\begin{equation}
\alpha_{i}(t)=\alpha_{i,0}\exp(-i\omega t)=\frac{b_{i}}{\omega^2_{0}-\omega^{2}-i\frac{\omega\omega_{0}}{Q}}\exp(-i\omega t).
\end{equation}

The power emission/loss of the cavity is related to the energy stored and the quality factor of the cavity as
\begin{equation}
\mathcal P_{\rm loss}=\frac{U}{Q}\omega_0.
\end{equation} 

Reasonably close to resonance ($|\omega^2-\omega^2_{0}|\ll\omega^{2}_{0}$) the energy stored inside the cavity is\begin{equation}
\label{energy}
U\approx\frac{|\alpha_{i,0}|^2 \omega^{2}_{0}}2\int d^3\mathbf{x}\, |\mathbf{A}^{\rm cav}_{i}(\mathbf{x})|^2. 
\end{equation}

This yields
\begin{equation}
\label{cavityloss}
\mathcal P_{\rm loss}=\chi^2 \rho_{\rm CDM} V\, \mathcal G\frac{\omega^{2}_{0}}{m_{\gamma^{\prime}}}\frac{1}{Q}\left|\frac{m^2}{\omega^2_{0}-\omega^{2}-i\frac{\omega\omega_{0}}{Q}}\right|^{2},
\end{equation}
where the geometric factor 
$\mathcal G$ is defined as
\begin{equation}
\mathcal G
= \frac{\left|\int dV{\mathbf{A}}^{*\rm cav}(\mathbf{x})\cdot \hat{\mathbf{n}}\right|^2}{V\, \int d^3\, \mathbf{x} |\mathbf{A}^{\rm cav}(\mathbf{x})|^2}.
\end{equation}

From Eq.~\eqref{cavityloss} it is quite clear that for large $Q$ emission from the resonant modes is clearly dominant.
While on resonance we have an enhancement by a factor $Q$, off-resonance we have a suppression by $1/Q$.
This is an enormous factor in particular for big cavities since then $Q\sim N_{\rm ref} \omega L_{\rm cavity}$ is particularly large even for moderate values of $N_{\rm ref}$.

\subsubsection{Large cavities}
With the general expressions for one cavity mode we can now proceed to calculate the total power emitted by the combination of all possible modes inside the cavity.
To be concrete and to keep the calculation explicit we consider the case of the cubic box cavity.

Before we start let us think about the physical situation. Since the dark matter particles are not at rest we actually have an energy distribution 
\begin{equation}
\label{distribution}
\rho_{\rm CDM}=\int^{\infty}_{m_{\gamma^{\prime}}}d\omega \hat{\rho}(\omega),
\end{equation}
with
\begin{equation}
\omega=\sqrt{m^2+p^{2}}\approx m^2\left[1+\frac{v^2}{2}+\ldots\right].
\end{equation}
As galactic matter is expected to have typical velocities of the order of $v\sim 10^{-3}$(=$300\, {\rm km}/{\rm s}$) the width of this distribution is approximately
\begin{equation}
\delta\omega \sim 10^{-6}\,m_{\gamma^{\prime}}.
\end{equation}

Let us now define very large cavities as those whose base frequency is smaller than the width of the DM distribution 
\begin{equation}
\label{many}
\wbase=\frac{\pi}{L_{\rm cavity}}\ll \delta\omega \quad \ (\rm large\, cavity). 
\end{equation}
The distance between two neighboring modes is typically smaller than the base frequency, 
\begin{equation}
\omega_{m+1,n,p}-\omega_{m,n,p}\lesssim \frac{\pi}{L_{\rm cavity}}=\wbase.
\end{equation}
from which it follows that in a large cavity we have a large number of resonance peaks within the energy distribution of the HPs. 

We have already mentioned that for the large cavities which we consider we expect very high $Q$ values typically $Q\gg 10^6$.
Therefore the width of the resonance peaks $\sim \omega_{0}/Q\sim \omega/Q$ is very small, much smaller than the width of the HP energy distribution
\begin{equation}
\label{narrow}
\frac{\omega_{0}}{Q}\ll \delta \omega.
\end{equation}

Equations~\eqref{many} and \eqref{narrow} mean that we have many narrow resonance peaks contributing to the total power output of the cavity. In this sense we have incoherent excitation of a large number of modes.

\subsubsection{Total power emitted by a box cavity}

Using the energy distribution of the HPs the power emitted into a single mode of the cavity is
\begin{equation}
\label{onemodecont}
P_{\rm 1-mode}=\chi^2V{\mathcal G}_{\rm mode}\int d\omega\hat{\rho}(\omega)\frac{\omega^{2}_{0}}{m_{\gamma^{\prime}}}\frac{1}{Q}\left|\frac{m^2}{\omega^{2}_{0}-\omega^2-i\frac{\omega\omega_{0}}{Q}}\right|^{2}.
\end{equation}
 
For large values of $Q$ the factor $\frac{1}{Q}\left|\frac{m^2}{\omega^2_{0}-\omega^{2}-i\frac{\omega\omega_{0}}{Q}}\right|^{2}$ approximates a $\delta$ function (times a factor).
Accordingly we obtain,
\begin{equation}
P_{\rm 1-mode}=\chi^2 V {\mathcal G}_{\rm mode} \frac{\pi}{2}\frac{m^3_{\gamma^{\prime}}}{\omega^{3} _{0}}\omega^{2}_{0}\hat{\rho}(\omega_{0}).
\end{equation}

To sum over all modes we now need the geometric factors. Let us consider a hidden photon oriented purely in the z-direction (generalization can be done analogously).
For our box cavity the relevant eigenmodes are those for the vector potential in the $z$-direction,
\begin{equation}
A_{z}=C_{n,m,p}\sin\left(\frac{m\pi x}{L}\right)\sin\left(\frac{n\pi y}{L}\right)\cos\left(\frac{p\pi z}{L}\right),
\end{equation}
and
\begin{equation}
\omega^2=\left(\frac{\pi}{L}\right)^2\left(m^2+n^2+p^2\right)=\wbase^2\left(m^2+n^2+p^2\right).
\end{equation} 
Inserting this one finds,
\begin{equation}
{\mathcal G}_{n,m,p}=\bigg\{\begin{array}{ccc}
\frac{64}{\pi^4}\frac{1}{n^2 m^{2}} & {\rm for} & n,m={\rm odd},\,\,p=0\\
0 & & {\rm otherwise}.
\end{array}
\end{equation}
Therefore we can from now on concentrate on the modes with $p=0$.
Moreover, it is clear that the dominant contribution will come from those modes where either $n$ or $m$ is as small as possible. Since the situation is symmetric in $m$ and $n$ it is sufficient to take one of them small and multiply the result by 2. 
Therefore let us concentrate on those modes with $(m,n)$ $m\gg n$. For those modes we have,
\begin{equation}
\omega_{m,n,}\approx m \,\wbase.
\end{equation} 
The distance between these types of modes is
\begin{equation}
\omega_{m+1,n,0}-\omega_{m,n,0}\approx \wbase.
\end{equation}

Let us now calculate the combined power from all modes within an interval $\Delta \omega$. Due to condition \eqref{many} we can pick this interval such that it is much smaller than the width of the HP energy distribution $\delta\omega\gg\Delta \omega$ allowing us to assume that the HP density $\hat{\rho}(\omega)$ is constant in this interval and at the same time make sure that there are 
still many suitable modes inside,
\begin{equation}
{\rm \#\,\,modes}\approx \frac{\Delta \omega}{2\wbase}\gg 1,
\end{equation}
where the factor of $2$ in the denominator arises because only odd modes contribute.

While the power from modes $(m,3)$, $(m,5)$ etc. is much smaller it is not entirely negligible. We can include it by summing over all those modes,
\begin{equation}
\sum_{n={\rm odd}}\frac{1}{n^2}=\frac{\pi^2}{8},
\end{equation}
where we have extended the sum to infinity which we can do due to the fast convergence of this series.

Combining everything so far we have,
\begin{equation}
P_{\rm interval}=2\frac{\Delta\omega}{2\wbase}\chi^2 V \frac{64}{\pi^4 m^2}\frac{\pi^2}{8} \frac{\pi}{2}\frac{m^{3}_{\gamma^{\prime}}}{\omega^3_{0}}\omega^{2}_{0}\hat{\rho}(\omega_{0})=4\chi^2 L^2 \hat{\rho}(\omega_{0})\Delta \omega 
\end{equation}
This can now be easily integrated to give the total output power from the cavity,
\begin{equation}
P^{\rm cubic}_{\rm mode\,\,expansion}=4 \chi^2 L^2 \int d\omega \hat{\rho}(\omega)=4\chi^2 L^2 \rho_{\rm CDM}.
\end{equation}
Comparing this with Eq.~\eqref{cubic1} we find that the two ways to calculate the emitted power do agree, as they should.

\subsection{From resonant to broadband}
Let us briefly consider also the transition region between the two limiting cases studied above.

As an example, let us take the same cubic cavity of length $L$. For the quality factor of the cavity we assume that it increases with the frequency as,
\begin{equation}
Q=Q_{0}\left(\frac{\omega}{{\rm wbase}}\right).
\end{equation}
Moreover, for concreteness we use a dark matter energy spectrum of width
\begin{equation}
\frac{\Delta \omega}{\omega}=\frac{1}{4Q_{0}}.
\end{equation}
This is chosen such that for the lowest frequency of the cavity the different dark matter particles all fit inside the resonance. 

\begin{figure}[t]
\begin{center}
\includegraphics[width=12cm]{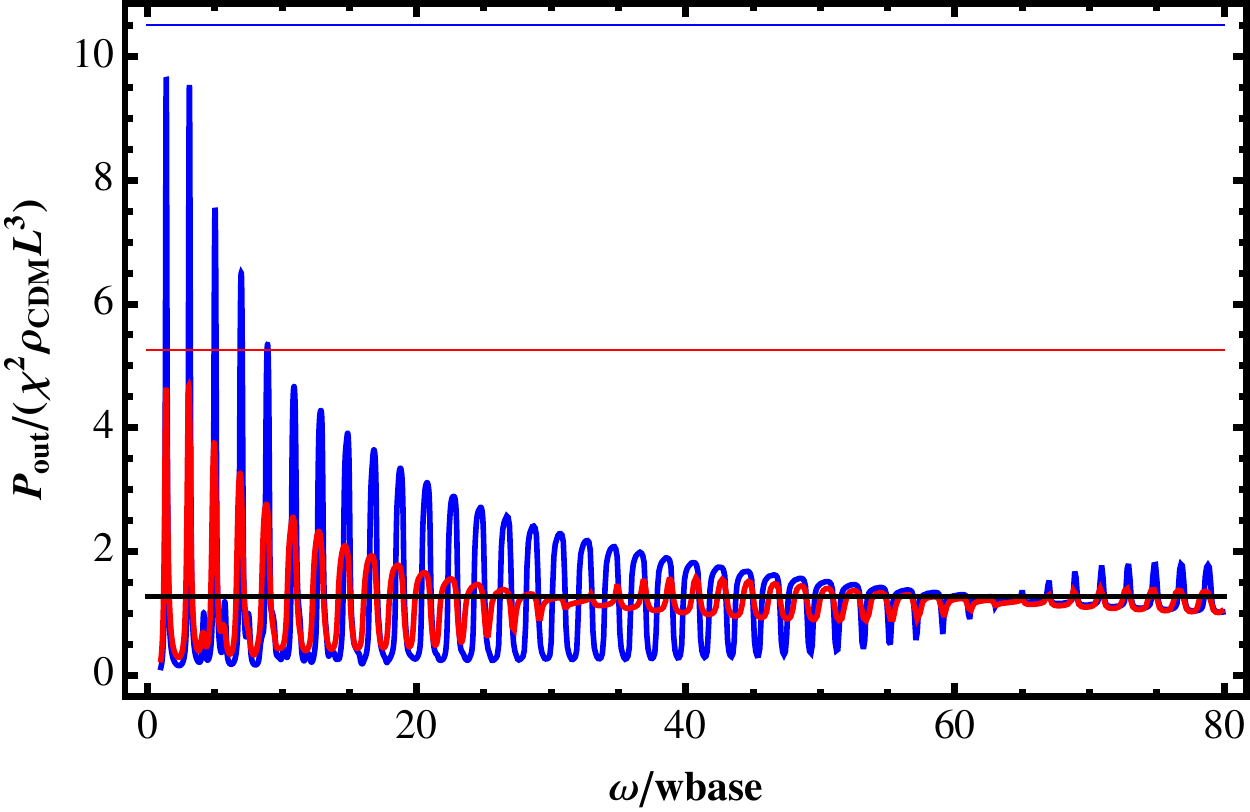}
\caption{Power output for a cubic cavity with length $L$ as a function of frequency. The blue curve corresponds to $Q_{0}=8$, the red one for $Q_{0}=4$.
On the left side we clearly see resonant enhancement by a factor $\sim Q$ of the mode. The thin blue and red lines denote the peak value expected for the lowest mode from Eq.~\eqref{cavityout}. The black line is the expectation in the $\omega\to\infty$ limit.}
\label{transition}
\end{center}
\end{figure}

We can now numerically sum over all modes, each contributing according to Eq.~\eqref{onemodecont}.
The results are shown in Fig.~\ref{transition} for two different value $Q_{0}=4,8$ (red and blue).

The lowest modes exhibit the highest resonant enhancements due to the largest geometry factors.
They nearly reach the approximate expectation from Eq.~\eqref{cavityout} (for the lowest mode this expectation is shown as the thin red and blue lines). The slight shortfall is due to the finite and pretty small $Q$ value, whereas Eq.~\eqref{cavityout} was derived for very large $Q$.
As the frequency increases both curves slowly approach the expectation of an area law, independent of the frequency (black line).
Comparing the two curves we see that the approach to the area law is slower for larger $Q_{0}$ (blue curve). This is in line with the expectation that higher $Q$ values lead to more pronounced resonant effects.

\section{Benefits of very large cavities for detection experiments?}\label{benefits}
Now that we have seen that very large cavities behave like broadband surface emitters let us see if this can be useful for a detection experiment. 

For a very large cavity we can imagine two different types of detector:
\begin{itemize}
\item[(i)]{} A localized detector with effective size much smaller than the cavity and absorbing only a small fraction of the power absorbed by the cavity walls.
\item[(ii)]{} A large detector covering, e.g., a large part of the inside surface of the cavity. This detector absorbes a sizeable fraction (ideally the same amount) as the cavity walls.
\end{itemize}

Let us start with (i).
In a very large cavity the detector may only occupy a small part of the total volume. For the detection it is therefore more sensible to consider the local energy density. For sufficiently incoherent emission it is also likely that the power will nicely distribute over the available volume. We can therefore consider the average energy density inside the cavity.
For this situation let us assume assume that the detector absorbs significantly less power than the surface. 
Using this we find that
\begin{equation}
\label{loss}
P_{\rm loss}=\omega\frac{E_{\rm cavity}}{Q},
\end{equation}
where $E_{\rm cavity}$ is the total energy inside the cavity and $Q$ is the quality factor.
Strictly speaking this equation holds for the energy loss in a given mode with quality factor $Q$. In our situation
with incoherent radiation many modes will be excited. However, for very high mode numbers the $Q$ factor typically does not depend very much on the specific mode (as we argue below it is basically determined by the reflectivity). Therefore we can take it as a typical $Q$ factor.

Since the power emitted from the surface must be equal to the power absorbed by the cavity walls (remember that we are neglecting the power absorbed by the detector) we can equate \eqref{loss}  and \eqref{surfaceemission} to determine the energy density inside the cavity,
\begin{equation}
\label{energydensity}
\epsilon_{\rm cavity}=\frac{E_{\rm cavity}}{V_{\rm cavity}}=\langle \cos^{2}(\theta) \rangle_{\rm surface}\chi^2\rho_{\rm CDM} 
Q\frac{A_{\rm cavity}}{\omega V}.
\end{equation}

From this formula it looks as if the energy density inside the cavity actually decreases with increasing volume. This is however not quite the case.
For very large mode numbers we can actually think of the light being reflected back and forth inside the cavity. 
Let us now consider a situation when the light is on average reflected $N_{\rm ref}$ times before being absorbed.
Then the typical photon spends a time $\sim N_{\rm ref}L_{\rm cavity}$ inside the cavity.
Accordingly we have
\begin{equation}
Q\sim \omega_{0} L_{\rm cavity} N_{\rm ref}.
\end{equation}
Inserting into Eq.~\eqref{energydensity} and using $A_{\rm cavity}\sim L^{2}_{\rm cavity}$ and $V_{\rm cavity}\sim L^{3}_{\rm cavity}$ we then have
\begin{equation}
\label{energydensity2}
\epsilon_{\rm cavity}\sim \langle \cos^2(\theta) \rangle_{\rm surface}\chi^2\rho_{\rm CDM} N_{\rm ref}.
\end{equation}

From Eq.~\eqref{energydensity2} we can see that while increasing the volume is neither beneficial nor detrimental the detectable energy density is increased by using a material with higher reflectivity.

In case (ii) we, by definition, absorb a power of similar order as that absorbed by the cavity walls. Accordingly we absorb a sizeable fraction of the total power generated by the cavity walls.
The power reaching the detector is therefore,
\begin{equation}
P_{\rm detector}\sim \langle \cos^{2}(\theta) \rangle_{\rm surface}\chi^2\rho_{\rm CDM} A_{\rm cavity}.
\end{equation}
This again increases with the size of the cavity.
However, for radiation evenly distributed inside the cavity we need a detector with a large effective area in order
to extract this power. This typically increases the background noise, e.g. thermal radiation.
Therefore, a special geometry that concentrates the generated power as, e.g. a dish antenna~\cite{Horns:2012jf} may be preferable.

\section{Conclusions}\label{conclusions}
In this note we have carefully studied the relation between resonant and broadband searches for WISPy dark matter.
Although different pictures (volume source vs. surface emission) are usually used to derive the signal we have argued that both are just two different cases of the same type of physics, as they should be.

Practically, for cavities which are very large compared to the Compton wavelength of the WISPy dark matter particles
the emitted power is essentially given by the surface area of the cavity. While a larger cavity therefore means more power generated, it is not trivial to detect this power because it is distributed over an even larger volume. Therefore, in the design of a broadband discovery experiment one should seek for ways to improve signal to noise, e.g. by concentrating the generated photons at the detector. 

\section*{Acknowledgements}

We'd like to thank the participants of the ``Dark Matter: a light move'' held in Hamburg DESY for 
very stimulating discussions that made this paper much easier to conceive and write. 
J.R.\ acknowledges support by the Alexander von Humboldt Foundation and 
partial support by the European Union through the Initial Training Network ``Invisibles,''

\end{document}